\documentclass{icrc2009}

\usepackage{graphicx}   
\usepackage{caption}    
\usepackage[font=footnotesize]{subfig} 
\usepackage{fixltx2e}
\usepackage{cite}
\usepackage{url}

\newcommand{\shorttitle}[1]%
{\markboth{Proceedings of the 31\MakeLowercase{$^{st}$} ICRC, {\L}\'{o}d\'{z} 2009}{#1} }
\newcommand{\etal}{\MakeLowercase{\textit{et al. }}} 

\begin{document}
\title{MAGIC observation of GRB\,080430}

\author{\IEEEauthorblockN{	S. Covino\IEEEauthorrefmark{1},
    M. Garczarczyk\IEEEauthorrefmark{2},
    M. Gaug\IEEEauthorrefmark{3},
    A. Antonelli\IEEEauthorrefmark{4},
    D. Bastieri\IEEEauthorrefmark{5},\\
    J. Becerra-Gonzalez\IEEEauthorrefmark{3},
    A. La Barbera\IEEEauthorrefmark{4},
    A. Carosi\IEEEauthorrefmark{4},
    N. Galante\IEEEauthorrefmark{6},
    F. Longo\IEEEauthorrefmark{7},\\
    V. Scapin\IEEEauthorrefmark{8},
    S. Spiro\IEEEauthorrefmark{4},
    A. de Ugarte-Postigo\IEEEauthorrefmark{9},
    A. Galli\IEEEauthorrefmark{10},
    R. Salvaterra\IEEEauthorrefmark{1}}
for the MAGIC collaboration \\
						\\

\IEEEauthorblockA{\IEEEauthorrefmark{1}INAF / Brera Astronomical Observatory, Via Bianchi 46, 23807, Merate (LC), Italy}
\IEEEauthorblockA{\IEEEauthorrefmark{2}IFAE, Edifici Cn, Campus UAB, 08193 Bellaterra, Spain}
\IEEEauthorblockA{\IEEEauthorrefmark{3}Instituto de Astrof\'isica de Canarias, via L\'actea s/n, 38205 La Laguna, Tenerife, Spain}
\IEEEauthorblockA{\IEEEauthorrefmark{4}INAF / Rome Astronomical Observatory, Via Frascati 33, 00044, Monte Porzio (Roma), Italy}
\IEEEauthorblockA{\IEEEauthorrefmark{5}Universit\`a di Padova and Istituto Nazionale di Fisica Nucleare (INFN), 35131, Padova, Italy}
\IEEEauthorblockA{\IEEEauthorrefmark{6}Max-Planck-Institut f\"ur Physik, F\"ohringer Ring 6, D-80805 M\"unchen, Germany}
\IEEEauthorblockA{\IEEEauthorrefmark{7}Dipartimento Fisica and INFN Trieste, 34127 Trieste, Italy}
\IEEEauthorblockA{\IEEEauthorrefmark{8}Universit\`a di Udine, and INFN Trieste, 33100 Udine, Italy}
\IEEEauthorblockA{\IEEEauthorrefmark{9}European Southern Observatory, Casilla 19001, Santiago 19, Chile}
\IEEEauthorblockA{\IEEEauthorrefmark{10}INAF / IASF, Via Fosso del Cavaliere 100, 00133 Roma, Italy}
}

\shorttitle{Covino \etal GRB\,080430 observation}
\maketitle

\begin{abstract}
Gamma-ray bursts are cosmological sources emitting radiation from the gamma-rays to the radio band. Substantial observational efforts have been devoted to the study of GRBs during the prompt phase, i.e. the initial burst of high-energy radiation, and during the longer-lasting afterglows. In spite of many successes in interpreting these phenomena there are still several open key questions about the fundamental emission processes, their energetics and the environment. Moreover, independently of their modeling, GRB spectra are remarkably simple, being satisfactorily fitted with power-laws, and therefore offer a very valuable tool to probe the extragalactic background light distribution affecting all high-energy observations of cosmological sources. Observations carried out with Cherenkov telescopes, as MAGIC, can be fundamental for all these scientific topics. GRB\,080430, being at a rather moderate redshift,  $z\sim0.76$, and well-studied in the optical, although observed only a few hours after the high-energy event, is a good test case to evaluate the perspective for late-afterglow observations with ground based GeV/TeV observatories.  
  \end{abstract}

 
\section{Introduction}
GRB\,080430 was detected by the \textit{Swift} satellite on April 30 2008 at 19:53:02\,UT \cite{Gui08}. 
An X-ray and optical counterpart were discovered and followed-up by many groups. Spectroscopy was rapidly carried out allowing to derive a redshift of 
$z = 0.758$ (the redshift estimate has been revised recently with a more accurate wavelength calibration \cite{deUg08,CuFo08,deUg09}). 
The relatively modest redshift made it an interesting target for MAGIC observations.
 
\section{MAGIC observations}

GRB080430 occurred while the Sun was still on the sky at the MAGIC site (Roque de los Muchachos, $28.75^\circ$N, $17.89^\circ$W). 
The MAGIC observation started immediately after sunset, at 21:12:14\,UTC and ended at 23:52:30\,UTC. The observation 
was disturbed by clouds so that 40\,min of the data, starting from 21:12:14\,UTC had to be rejected. 
The begin of the observation was at T$_0$ + 4753\,s, well after the end of the prompt emission phase. 
The observation with MAGIC started at a zenith angle of Z$_{\rm d} = 23^\circ$, reaching Z$_{\rm d} = 35^\circ$ at the end.  
Analysis of the dataset, in the energy bin from 80 up to 125\,GeV, gave a 95\% CL upper limit for the emission from the source 
of $F_{\rm 95\%\,CL} = 5.5 \times 10^{-11}$\,erg\,cm$^{-2}$\,s$^{-1}$ (under the assumption of steady emission) or a fluence limit 
of F$_{\rm 95\%\,CL} = 3.5 \times 10^{-7}$\,erg\,cm$^{-2}$ for a time interval of 6258\,s from 21:12:14~UT to 22:56:32~UT. 
These limits contain a 30\% systematic uncertainty on the absolute detector efficiency. 
Limits at higher energies are less important for the present analysis due to intense Extragalactic Background Light (EBL) absorption above $\sim 100$\,GeV 
(see Sect.\,\ref{sec:ebl}). It is important to note that at that time, the sum trigger hardware upgrade~\cite{Alb08,GRB} was not yet available for GRB observations
 and therefore the lowest obtained upper limit is a factor two higher than in later cases~\cite{GRB090102}.

\section{Afterglow light-curve and spectral energy distribution}

It is not our purpose to discuss in detail the physics of the afterglow of this event which will be discussed by~\cite{deUg09}. 
At the epoch of the MAGIC observations (about 8\,ksec from the high-energy event), the afterglow seems to have entered a rather stable phase 
with no big flares although possibly with small bumps affecting the afterglow evolution. As it happens frequently, no standard afterglow model 
really fits perfectly the observations, and there is still the possibility of a more complex description with more components playing a role
 (energy injection, structured jets, etc.). However, most of these details do not essentially affect our analysis. 
Data seem to be more consistent with a constant circumburst density environment with electron distribution index $p \sim 2.0$ or slightly more
 if some energy injection is going on as it could be the case. From the modeling of the SED with a limited amount of rest frame absorption the
 synchrotron cooling frequency seems to be located somewhere between the optical and X-ray bands. 

From these inferences, we can derive an electron distribution index $p$ and choose which environment allows us to better model the data. 
Locating when possible, or computing from the models, the typical synchrotron and cooling frequency location, we can also derive the micro-physical
 parameters  $\epsilon_{\rm e}$, the fraction of energy going to electrons, and  $\epsilon_{\rm B}$, the fraction of total energy going to magnetic fields.

\subsection{Total energy}

The total energy can be derived from the burst isotropic energy E$_{\rm iso}$ with some assumptions and correcting it for the
fireball radiative efficiency $\eta$. We estimate E$_{\rm iso}$ as the integral of the burst spectral model in the $1 - 10^4$\,keV band~\cite{Ama02}. 
In this energy band the spectrum of a burst is typically described by a Band model, and in order to calculate this integral we need to know the two
power-law photon indices and the peak energy E$_{\rm peak}$. We retrieved from the on-line \textit{Swift} GRB table the $15 - 150$\,keV fluence of the burst
and its photon index. 
We therefore run a set of integrations by varying E$_{\rm peak}$ and derive the corresponding E$_{\rm iso}$ using the $15 - 150$\,keV BAT fluence to normalize the spectrum. 
The chosen values of E$_{\rm peak}$ and E$_{\rm iso}$ are those satisfying the Amati relation~\cite{Ama02}. 
In each integration, depending on the value of E$_{\rm peak}$, we identify the observed photon 
index with one of the two indices of the Band function, fixing the other one to a canonical value
(1 and 2 for the low- and the high-energy power-laws, respectively). According to this method, we estimate E$_{\rm peak} = 39$\,keV, 
and E$_{\rm iso} = 3 \times 10^{51}$\,erg. 
If we then assume a radiative efficiency $\eta$ of 10\%, then we have E$_{\rm iso} = 3 \times 10^{52}$\,erg. 
The relatively low peak energy can allow one to classify this event as X-Ray Flash or X-Ray Rich. 
The derived total energy is on the average for cosmological GRBs, although it is not uncommon to observe events substantially more energetic~\cite{Sak08}.

Summing up, we can assume to know, with various degrees of uncertainties, all the required parameters for a modeling of the high energy emission 
from the afterglow of GRB\,080430: energy  E$_{\rm iso} \sim 3 \times 10^{52}$\,erg, $\epsilon_{\rm e} \sim 0.1$, $\epsilon_{\rm B} \sim 0.01$ 
(not really constrained by the observations, but in agreement with them), $p \sim 2.3$ (assuming some degree of energy injection), the circumburst 
medium density profile $n \sim 1$\,cm$^{-3}$ and the redshift $z \sim 0.76$. Our observations are at $t \sim 8$\,ksec from the burst and the afterglow 
synchrotron emission is in the so called "slow-cooling" regime (i.e. the synchrotron cooling frequency is above the synchrotron typical frequency). 

\section{Self-Synchrotron Compton during the afterglow}

The analysis of the high energy emission from the various phases of a GRB has been considered by many authors (see~\cite{Cov09} for a recent review). 
In our case, we can restrict us to essentially one only process, Self-Synchtron Compton (SSC). In fact, we performed observations with a delay from the 
GRB onset long enough (about two hours) to rule out any residual prompt emission and, moreover, the light-curve does not show any dominant flaring 
at the observation epoch. This means that we are in a rather simple situation, and once the parameters of the lower-energy synchrotron emission are known, 
it is possible to predict the SSC component with good reliability.

Among the many possible sources, we followed the recipe described by~\cite{FaPi08}.

One of the most intriguing features of the SSC process, is that it essentially generates a new spectral component superposed to the underlying synchrotron 
spectrum but with the same global shape up to a cut-off frequency:

\begin{equation}
\nu_{\rm M,SSC} \sim \frac{\Gamma^2 m_{\rm e}^2 c^4}{h^2 \nu_{\rm c}},
\end{equation}

where $\nu_{\rm c}$ is the synchrotron cooling frequency, $\Gamma$ is the fireball bulk motion Lorentz factor, $m_{\rm e}$ the electron mass, $c$ the speed of light 
and $h$ the Planck constant. Above this frequency the SSC emission is no more in the Thompson regime and becomes much weaker (Klein-Nishina regime). 
For typical parameters the SSC emission of the afterglow is however in the Thomson regime.

Assuming we are in a constant density circumburst environment, the predicted SSC spectrum is characterized by two typical frequencies:

\begin{eqnarray}
\nu_{\rm m,SSC} \approx & 6.2 \times 10^{21}\,{\rm Hz}~C_p^4\,\epsilon_{{\rm e},-1}^4\,\epsilon_{{\rm B},-2}^{1/2}\,n^{-1/4}  \nonumber \\
& E_{\rm k,53}^{3/4}\,t_3^{-9/4}\,(1+z)^{5/4},
\label{eq:mssc}
\end{eqnarray}

\begin{eqnarray}
\nu_{\rm c,SSC} \approx & 4 \times 10^{24}\,{\rm Hz}~(1+Y_{\rm SSC})^{-4}\,\epsilon_{{\rm B},-2}^{-7/2}\,n^{-9/4} \nonumber \\
& E_{\rm k,53}^{-5/4}\,t_3^{-1/4}\,(1+z)^{-3/4},
\label{eq:vssc}
\end{eqnarray}

where $C_p = 13(p-2)/\left[3(p-1)\right]$ and $Y_{\rm SSC} = U'_{\rm syn}/U'_{\rm B}$ is the rest frame synchrotron to magnetic field energy density ratio 
and $z$ the source redshift. 

Defining $\eta$ as:

\begin{equation}
\eta = (\nu_{\rm m}/\nu_{\rm c})^{(p-2)/2}.
\end{equation}

It can be shown that\footnote{Here we deliberately ignore the possibility to have higher order IC components which could be effective in cooling the electron population.}:

\begin{equation}
Y_{\rm SSC} \simeq \frac{-1+\sqrt{1+4\eta\epsilon_{\rm e}/\epsilon_{\rm B}}}{2}.
\end{equation}

The peak synchrotron frequency to cooling synchrotron frequency ratio is:

\begin{equation}
\nu_{m} / \nu_{c} \simeq 0.024\,(1+z)\,C_p^2\,\epsilon_{{\rm e},-1}^2\,\epsilon_{{\rm B},-2}^2\,n\,E_{\rm k,53}\,t_3^{-1}.
\label{eq:nmncr}
\end{equation}

With our parameters $C_p = 1.0$ and Eq.\,(\ref{eq:nmncr}) becomes $ \nu_{m} / \nu_{c} \simeq 0.0016 $ and $Y_{\rm SSC} \simeq 1.51$. Eqs.\,(\ref{eq:mssc}) and (\ref{eq:vssc}) become $\nu_{\rm m,SSC} \approx 4.8 \times 10^{19}$\,Hz ($\simeq 200$\,KeV)  and $\nu_{\rm c,SSC} \approx 1.8 \times 10^{23}$\,Hz ($\simeq 730$\,MeV). The energy band covered by the MAGIC observations (E$_{\rm MAGIC} \sim 90$\,GeV) are clearly much higher than the cooling SSC frequency. We are therefore in the spectral range where the SSC spectrum is softer, following a power-law behavior, $\nu^{-p/2} \simeq \nu^{-1.15}$.  

In order to derive the expected flux density at the MAGIC observation energy we have to compute the flux density at the typical SSC frequency:

\begin{eqnarray}
F_{\nu_{\rm m},{\rm SSC}} \simeq & 10^{-11}\,{\rm erg\,cm}^{-2}\,{\rm s}^{-1}\,{\rm MeV}^{-1}~0.07\,n^{5/4} \nonumber \\
& \epsilon_{{\rm B},-2}^{1/2}\,E_{\rm k,53}^{5/4}\,t_3^{1/4}\,\left(\frac{1+z}{2}\right)^{3/4}\,D_{\rm L,28.34}^{-2},
\end{eqnarray}

where $D_{\rm L}$ is the luminosity distance of the source, $D_{\rm L} \sim 4.8$\,Gpc ($\sim 1.5 \times 10^{28}$\,cm). With our parameters, $F_{\nu_{\rm m},{\rm SSC}} \simeq 5.2 \times 10^{-13}$\,erg\,cm$^{-2}$\,s$^{-1}$\,MeV$^{-1}$.

Then, finally, from the peak energy to the MAGIC band we have to extrapolate the SSC spectrum as:

\begin{equation}
F_{90\,{\rm GeV}} \sim F_{\nu_{\rm m,SSC}} (\frac{\nu_{\rm c,SSC}}{\nu_{\rm m,SSC}})^{-(p-1)/2}\,(\frac{\nu}{\nu_{\rm c,SSC}})^{-p/2},
\label{eq:extrap}
\end{equation}

and, again with our parameters, $F_{90\,{\rm GeV}} \sim 9.8 \times 10^{-18}$\,erg\,cm$^{-2}$\,s$^{-1}$\,MeV$^{-1}$. The flux integrated in the MAGIC band, 
the parameter to be compared to the reported upper limits, can be well approximated by $\nu F_\nu$ at about 90\,GeV, 
and we have $F_{\rm MAGIC}  \sim 8.8 \times 10^{-13}$\,erg\,cm$^{-2}$\,s$^{-1}$.

\begin{figure}[!t]
 \centering
 \includegraphics[width=\columnwidth]{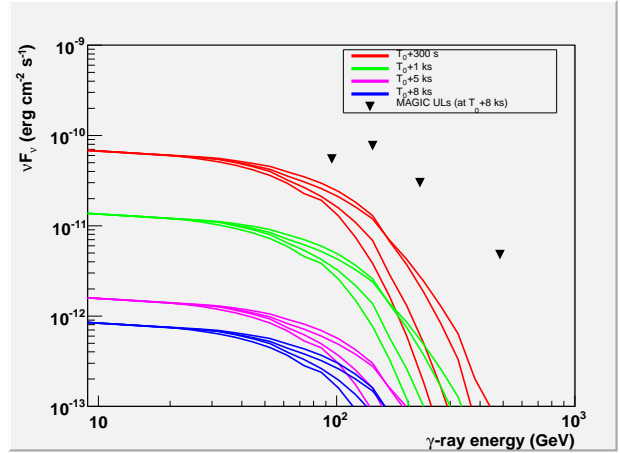}
 \caption{Predictions at different time delays from the high-energy event for the SSC emission during the afterglow of GRB\,080430. 
Black triangles are 95\% CL upper limits derived by MAGIC at various energies. Lines of a same color show the same SSC model, but a different absorption model of the gamma-rays by the EBL. The blue lines correspond to the MAGIC observation window.}
 \label{fig:ul}
\end{figure}

\section{Extragalactic background light attenuation}
\label{sec:ebl}

Gamma-rays in the GeV energy regime are absorbed through scattering processes with the Extra-Galactic Background Light (EBL). 
The precise light content of the EBL is strongly debated. In principle we can rely on many different models the predictions of which at $z \sim 1$ 
span a wide range of optical depths, from about 1 up to 6 \cite{FaPi08}. Recently, the MAGIC collaboration published a striking observational result~\cite{Alb08} 
suggesting the EBL attenuation could be much lower than previously assumed. This means that at the redshift of GRB\,080430, $z \sim 0.76$, 
and at the MAGIC energy, E $\sim 90$\,GeV, an optical depth, $\tau$, not far from unity is possible, however predictions of the strength of absorption 
can vary by up to a factor 6, at the distance of GRB080430. 
We included four representative models from~\cite{Fran08,Gilm09} and~\cite{Knei04}  and show the range of possible  absorbed spectra in Figure~\ref{fig:ul}.
The blue lines correspond to the MAGIC observation delay, the other lines show the spectrum at earlier observation times, in principlie easily accessible to IACTs. 

On average, we can derive an attenuation of the received flux from the afterglow of GRB\,080430 of the order a factor 3 or even slightly less, 
allowing us to estimate $F_{\rm MAGIC}  \sim 3 \times 10^{-13}$\,erg\,cm$^{-2}$\,s$^{-1}$ as the predicted flux in the MAGIC band.

\section{Conclusions}

The prediction of the expected SSC flux for an afterglow is not straightforward since it is required to know, or at least to reliably estimate, 
the parameters of the underlying afterglow (see Fig.\,\ref{fig:ul}). In the case of GRB\,080430 the sampling of the X-ray and optical afterglow is very good 
and we could constrain the various afterglow parameters in a sufficient way to derive meaningful predictions for the expected SSC flux.

The obtained results appear to be well below the reported upper limits. Moreover, even if we did not always use detailed afterglow parameters (not always available), but sometimes ``generic'' afterglow parameters, it is very unlikely  the predicted flux changes substantially. We also assumed a low opacity due to the EBL, this is not in disagreement with the present observations but the discussion is still very active in the field. Nevertheless, this pilot case shows fairly interesting perspectives for a late-afterglow detection at high energies. 

In general, to improve the expected flux from a GRB afterglow (for SSC) it is mandatory to try to decrease the observation energy (due to the $\nu^{-p/2}$ dependence above the cooling SSC frequency), which is also very important for the minimization the effect of the EBL attenuation too. If the telescope sum trigger hardware upgrade had already been finished before the observations, a limit above an energy of 45\,GeV would have been obtained (see also~\cite{GRB090102}). At these energies, the strong effect of the EBL could probably be neglected, apart from a factor 6 higher intrinsic flux in $\nu F_{\nu}$. The low energy threshold is likely the most important factor and the expected performances of MAGIC\,II will undoubtedly increase the chances of positive detections. 

As a matter of fact, GRB\,080430 was a rather average event in terms of energetics. More energetic GRBs are indeed relatively common, and due to the mild positive dependence on the isotropic energy of a GRB, much higher fluxes than in the present case can be allowed.

The time delay of the observation from the GRB has a clear impact essentially because the observed SSC component is strictly related to the underlying synchrotron component which rapidly decays in intensity with time, depending on the specific environment and micro-physical parameters. Eq.~\ref{eq:extrap} goes roughly with $t^{-1.34}$ which means that had MAGIC been able to start observations right at the start of the late afterglow phase (e.g. at $T_0 + 1$ks), the flux predictions had increased by a factor 8. 
The time delay for these observations of about two hours, coupled with the mediocre observing conditions, were more than enough to depress the observed flux and raise the reported upper limits. 

The case of GRB\,080403 demonstrates that if three conditions are met altogether: 1. a moderate redshift, 2. start of observations right at the beginning of the afterglow phase and 3. the usage of the MAGIC sum trigger providing energy thresholds below 50~GeV, detections of the IC component of the afterglow of GRBs are within reach. 

\subsection*{Acknowledgements}
We thank Yizhong Fan for useful discussions.
The collaboration thanks the Instituto de Astrof\'ısica de Canarias for the excellent working conditions at the Observatorio
del Roque de los Muchachos in La Palma, as well as the German BMBF and MPG, the Italian INFN and Spanish MCINN. 
This work was also supported by ETH Research Grant TH 34/043, by the Polish MniSzW Grant N N203 390834, and by the YIP of the Helmholtz
Gemeinschaft.


\begin{thebibliography}{999999}
\bibitem{Alb08} Albert J., Aliu E., Anderhub H. et al. 2008, Sci, 320, 1752
\bibitem{Ama02} Amati L., Frontera F., Tavani M. et al. 2002, A\&A 390, 81
\bibitem{Cov09} Covino S., Garczarczyk G., Galante N. et al. 2009, astro-ph/0901.1589
\bibitem{CuFo08} Cucchiara A., \& Fox D.B. 2008, GCN\,7654
\bibitem{deUg08} de Ugarte Postigo A., Christensen L., Gorosabel J. et al. 2008, GCN\,7650
\bibitem{deUg09} de Ugarte Postigo A., et al. 2009, in preparation
\bibitem{GRB} Garczarczyk M. et al., ''GRB Observations with the MAGIC Telescope'', these proceedings
\bibitem{GRB090102} Gaug M. et al., ''Observation of GRB090102 with the MAGIC Telescope'', these proceedings
\bibitem{FaPi08} Fan Y-Z., Piran T. 2008, Frontiers of Physics in China, 3, 306
\bibitem{Fran08} Franceschini et al., 2008, arXiv/0805.1841v1
\bibitem{Gilm09} Gilmore et al., 2009, arXiv:0905.1144v1
\bibitem{Knei04} Kneiske et al. 2004, A\&A 413, 807 
\bibitem{Gui08} Guidorzi C., Barthelmy S.D., Beardmore A.P. et al. 2008, GCN\,7647
\bibitem{Sak08} Sakamoto, T., Hullinger D., Sato, G. et al. 2008, ApJ, 679, 570
\end{thebibliography}
\end{document}